\documentclass[epj]{svjour}
\usepackage{latexsym}
\usepackage{graphics}
\usepackage{epsfig}
\usepackage{graphicx}
\usepackage{amsmath}
\usepackage{wrapfig}
\begin{document}
\title{Centrality dependence of freeze-out temperature fluctuations in Pb-Pb collisions at the LHC}
\author{Dariusz Prorok
\thanks{e-mail: dariusz.prorok@ift.uni.wroc.pl}%
}                     
%
%
\institute{Institute of Theoretical Physics, University of
Wroc{\l}aw, Pl.Maksa Borna 9, 50-204  Wroc{\l}aw, Poland}
\date{Received: date / Revised version: date}
%
\abstract{
Many data in the High Energy Physics are, in fact, sample means. It is shown that when this exact meaning of the data is taken into account and the most weakly bound states are removed from the hadron resonance gas, the whole spectra of pions, kaons and protons measured at midrapidity in Pb-Pb collisions at $\sqrt{s_{NN}} = 2.76$ TeV can be fitted simultaneously. The invariant distributions are predicted with the help of the single-freeze-out model in the chemical equilibrium framework. The method is applied to the measurements in centrality bins of Pb-Pb collisions and gives acceptable fits for all but peripheral bins. The comparison with the results obtained in the framework of the original single-freeze-out model is also presented. Some more general, possible implications of this approach are pointed out.
} 
\maketitle
\section{Introduction}
\label{intro}

The unprecedented success of physics in modern times is the result of the application of two general principles: the theoretical modeling of a phenomenon and the experimental verification (of the predictions of the model). One of the currently most explored part of the standard model is the theory of strong interactions - the Quantum Chromodynamics (QCD). The QCD predicts a transition from a system of hadrons (strongly interacting particles which can be observed) to a system of partons (quarks and gluons which cannot be observed individually). This requires extremely high temperatures or densities of the system. The conditions necessary for the appearance of the deconfined phase (the partonic system) of QCD can be established in the laboratory now (for a wide review of the subject, from the theory to the experiment, see Ref.~\cite{Leupold:2011zz}).

High-energy heavy-ion collisions are the tools for the creation of the deconfined phase. The matter originated during such a collision, extremely dense and hot, is compressed more or less in the volume of the narrow disc of the ion radius at the initial moment. After then the matter rapidly expands due to the tremendous pressure and cools simultaneously. The evolution of the matter can be described in the framework of the relativistic hydrodynamics \cite{Huovinen:2013wma}. During expansion the matter undergoes a transition to a hadron gas phase. The hadron gas continues the hydrodynamical evolution, assuming that the collective behavior does not cease at the transition. The expansion makes the gas more and more diluted, so when mean-free paths of its constituents become comparable to the size of the system one cannot treat the gas as a collective system. This moment is called \textit{freeze-out}. After then the gas disintegrates into freely streaming particles which can be detected. In principle, one can distinguish two kinds of freeze-out: \textit{a chemical freeze-out}, when all inelastic interactions disappear  and \textit{a kinetic freeze-out} (at lower temperature), when also elastic interactions disappear. The measured hadron yields are fingerprints of corresponding hadron abundances present at the chemical freeze-out. The yields can be consistently described within the grand canonical ensemble with only three independent parameters, the chemical freeze-out temperature $T_{ch}$, the baryochemical potential $\mu_B$ and the volume of the system at the freeze-out, $V$ \cite{BraunMunzinger:2003zd}. This idea is the fundament of the Statistical Model (SM) of particle production in heavy-ion collisions. The measured $p_T$ spectra include information about the transverse expansion (radial flow) of the hadron gas and the temperature $T_{kin}$ at the kinetic freeze-out \cite{Heinz:2004qz}. However, the alternative approach to freeze-out was founded in \cite{Broniowski:2001we,Broniowski:2001uk} where the single freeze-out was postulated, i.e. the kinetic freeze-out coincided with the chemical freeze-out. This is the Single-Freeze-Out Model (SFOM). The suitably chosen freeze-out hypersurface and the complete inclusion of contributions from resonance decays enabled to correctly describe the Relativistic Heavy Ion Collider (RHIC) $p_T$ spectra.

With the first data on Pb-Pb collisions at $\sqrt{s_{NN}} = 2.76$ TeV from CERN Large Hadron Collider (LHC) \cite{Abelev:2012wca,Abelev:2013vea} two new problems have appeared when the SM and hydrodynamics were applied for the description of particle production. The predicted proton and antiproton abundances were larger then measured ones \cite{Stachel:2013zma} and low $p_{T}$ pions were underestimated \cite{Abelev:2012wca,Melo:2015wpa}. This caused that the ratio $p/\pi = (p + \bar{p})/(\pi^{+} + \pi^{-})$ was overestimated in the SM by a factor $\sim 1.5$ \cite{Floris:2014pta}. Various explanation of this ''puzzle'' have been invented, but all fall outside the SM. These are: (i) the incomplete list of resonances, there could still be undiscovered (high mass) resonances which after decays would increase more pion yields than proton ones, (ii) the non-equilibrium thermal model, with two additional parameters describing the degree of deviation from the equilibrium, (iii) hadronic inelastic interaction after hadronization and before chemical freeze-out, especially baryon annihilation, and (iv) flavor hierarchy at freeze-out, which could result in two different freeze-out temperatures, one for non-strange hadrons, another for strange hadrons (for more details and references see \cite{Floris:2014pta}). And the later one: (v) inclusion of resonance spectral functions \cite{Huovinen:2016xxq,Vovchenko:2018fmh}.

In this work the generalization of the SFOM in the chemical equilibrium framework is postulated, which proved to be successful in the solution of the above  problems \cite{Prorok:2015vxa} and well reproduces the results of \cite{Abelev:2013vea}. This approach might be consider as the alternative (the (vi)$th$ ) possibility to the five ones listed above. However, in opposite to the original version of the SFOM, all parameters of the model (thermal and geometric) are estimated simultaneously from the spectra. This version was successfully applied to the description of the final spectra measured at RHIC for all centrality classes in the broad range of collision energy \cite{Prorok:2007xp}. The new idea introduced into the SFOM in the present work is to randomize one of the parameters of the model. The model will be called the Randomized Single-Freeze-Out Model (RSFOM) from now on. It has turned out that the successful improvement is achieved only when the freeze-out temperature becomes a random variable and nothing is gained with the randomization of geometric parameters of the model. This approach was applied successfully to the most central bin of Pb-Pb collisions at $\sqrt{s_{NN}} = 2.76$ TeV in \cite{Prorok:2015vxa}, for instance the ratio $p/\pi$ was explained. In the present paper results for all centrality classes of the above-mentioned collisions are reported.


\section{The model}
\label{modeldescr}

To the favour of the reader we hereby repeat the description of the method, which
is the same that was used in \cite{Prorok:2015vxa}.

In the SFOM the invariant distribution of the measured particles of species $i$ has the form

\begin{equation}
{ \frac{dN_{i}}{d^{2}p_{T}\;dy} }=\int
p^{\mu}d\sigma_{\mu}\;f_{i}(p \cdot u) \;, \label{Cooper}
\end{equation}

\noindent where $d\sigma_{\mu}$ is the normal vector on a
freeze-out hypersurface, $u_{\mu}=x_{\mu}/\tau_f$
is the four-velocity of a fluid element and $f_{i}$ is the final
momentum distribution of the particle in question. The final
distribution means that $f_{i}$ is the sum of primordial and
decay contributions to the distribution. The freeze-out hypersurface is defined by the equations

\begin{equation}
\tau_f = \sqrt{t^2-x^2-y^2-z^2}\;,\;\;\;\sqrt{x^2+y^2} \leq \rho_{max} \;,
\label{Hypsur}
\end{equation}

\noindent where the invariant time, $\tau_f$, and the transverse size, $\rho_{max}$, are two geometric parameters of the model. For the LHC energies all chemical potentials can be put equal to zero, so the freeze-out temperature, $T_f$, is the only thermal parameter of the model. The contribution from the weak decays concerns (anti-)protons mostly \cite{Abelev:2013vea,Milano:2012eea}, hence secondary (anti-)protons from primordial and decay $\Lambda$($\bar{\Lambda}$)'s are subtracted.

\begin{figure}
\resizebox{0.46\textwidth}{!}{%
 \includegraphics{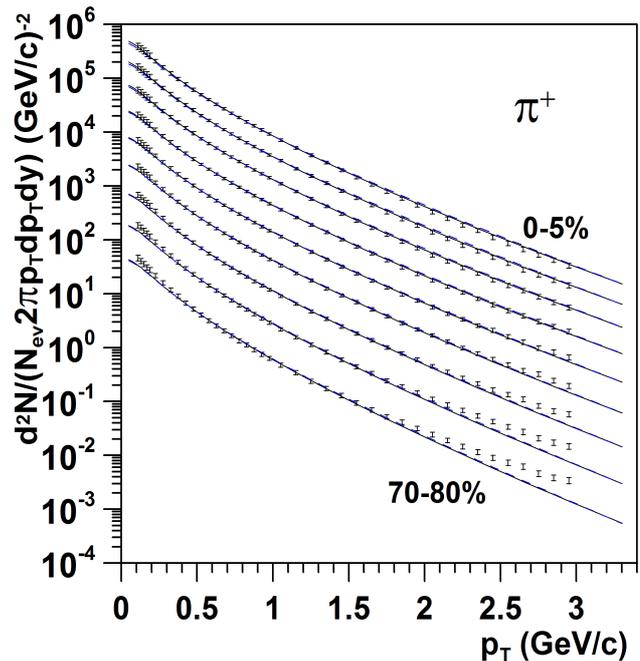}
}
\caption{Spectra of positive pions measured in Pb-Pb collisions at $\sqrt{s_{NN}} = 2.76$ TeV, data used in the fit are presented as error bars only \protect\cite{Abelev:2013vea}, errors are sums of statistical and systematic components added in quadrature. Central to peripheral data are shown, spectra are scaled by factors $2^n$ (peripheral data not scaled). Lines are fits for the log-normal p.d.f. of $\beta_f$, dashed lines (blue) show fits of the SFOM without randomization and with all hadronic resonances included. }
\label{fig:1}
\end{figure}

\begin{figure}
\resizebox{0.46\textwidth}{!}{%
 \includegraphics{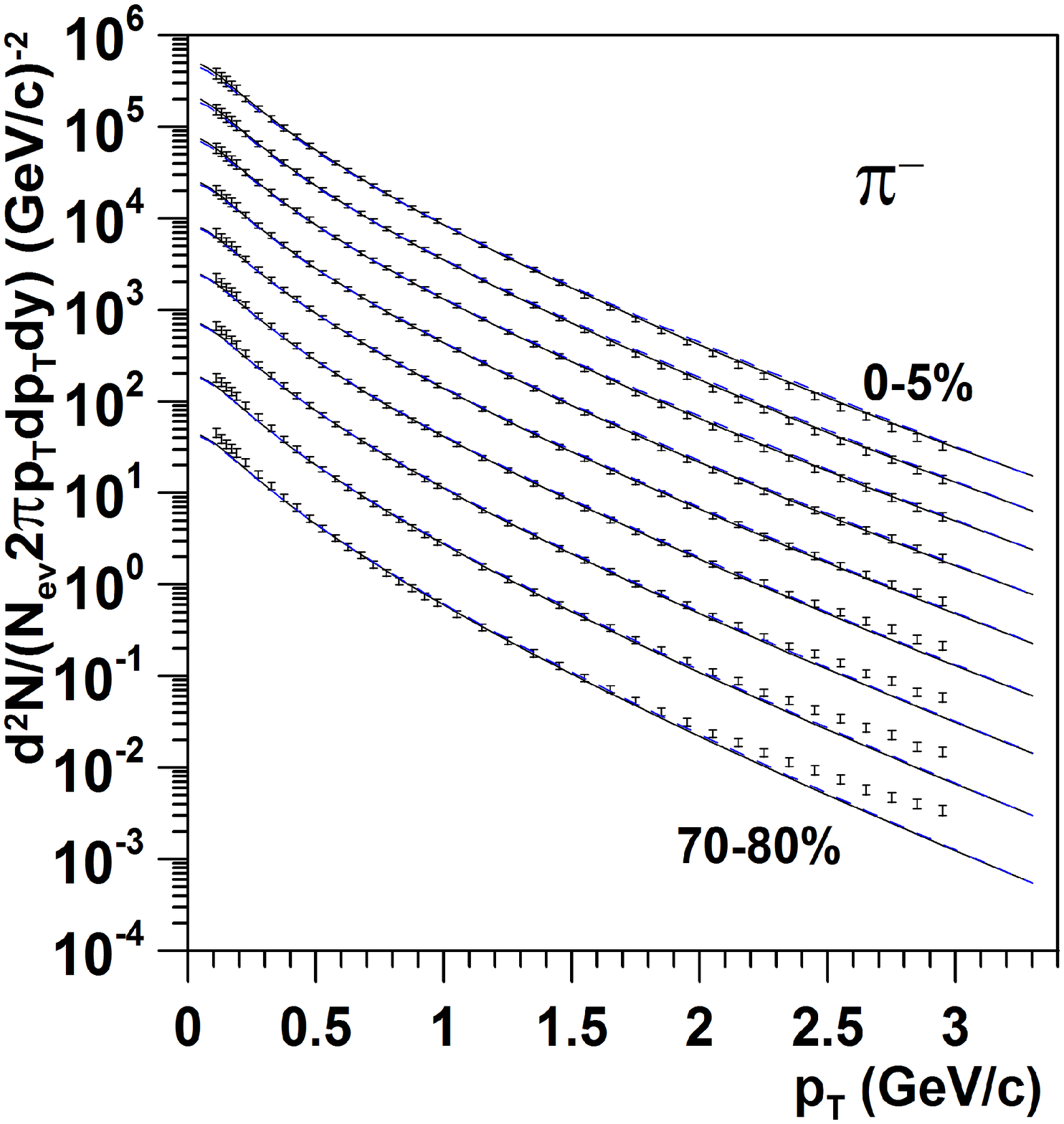}
}
\caption{The same as Fig.\,\ref{fig:1} but for negative pions. }
\label{fig:2}
\end{figure}

However, the data on $p_{T}$ spectra \cite{Abelev:2012wca,Abelev:2013vea} are not 'points' but, what is called in statistics, \textit{sample means} (the division by $N_{ev}$ - the number of events in the sample, means that \footnote{the sample is a centrality class here.}). In the large sample limit (the sample size goes to infinity), a sample mean converges to a distribution (theoretical) mean, not to just one value of the theoretical equivalent of a measurand (here Eq.~(\ref{Cooper})). This is guaranteed by \textit{the weak law of the large numbers} \cite{Cowan:1998ji,James:2006zz}. Therefore, the theoretical prediction should be also a random variable and the quantity to compare with the data - its average. For simplicity it is assumed that the theoretical prediction, Eq.~(\ref{Cooper}), is a statistic (a function of a random variable, by definition it is also a random variable) and that one of the parameters of the model, $\theta$ ($\theta = T_f, \tau_f \; \textrm{or}\; \rho_{max}$), is a random variable. Then the theoretical prediction becomes the appropriate average:
\begin{equation}
\left\langle { \frac{dN_{i}}{d^{2}p_{T}\;dy} } \right\rangle_{\theta} = \int { \frac{dN_{i}}{d^{2}p_{T}\;dy} } f(\theta) d\theta \;, \label{AvCoop}
\end{equation}
\noindent where $f(\theta)$ is the probability density function (p.d.f.) of $\theta$. This approach is more general but includes the standard one, if fluctuations of $\theta$ are negligible, then its p.d.f. is Dirac-delta like, $f(\theta) \sim \delta(\theta-\theta_o)$ and the average becomes the value at the optimal point $\theta_o$. It has turned out that only randomization of $T_f$ improves the quality of the fit, randomization of $\rho_{max}$ or $\tau_f$ does not change anything. In fact, for the technical reasons, not $T_f$ is randomized but $\beta_f = 1/T_f$. From the statistical point of view these two possibilities are equivalent, because $\beta_f(T_f)$ has a unique inverse and vice versa \cite{Cowan:1998ji}. Two p.d.f.'s are considered: log-normal
\begin{equation}
f(\beta_f;\mu,\sigma) = \frac{1}{\sqrt{2\pi}\sigma}\frac{1}{\beta_f}
\exp \left\{-\frac{(\ln{\beta_f}-\mu)^2}{2\sigma^2} \right\} \; \label{lognorm}
\end{equation}
\noindent and triangular
\begin{equation}
f(\beta_f;\breve{\beta}_f,\Gamma) = \begin{cases} \frac{\Gamma - \mid \beta_f-\breve{\beta}_f \mid}{\Gamma^2}\;, \mid \beta_f-\breve{\beta}_f \mid \leq \Gamma  \cr  0\;\;\;\;\;\;\;\;\;\;\;\;\;\;\;, \mid \beta_f-\breve{\beta}_f \mid > \Gamma \;. \end{cases}  \label{triang}
\end{equation}
\noindent where $\mu$ and $\sigma$ are parameters of the log-normal p.d.f. whereas $\breve{\beta}_f$ and $\Gamma$ are parameters of the triangular p.d.f., $\breve{\beta}_f$ is the average of $\beta_f$. The first is differentiable but has an infinite tail, the second is not differentiable but has a finite range. The choice is arbitrary, but two general conditions should be fulfilled, a p.d.f. is defined for a positive real variable and has two parameters so as the average and the variance can be determined independently.

However, in both cases of p.d.f.'s, Eqs.~(\ref{lognorm}) and (\ref{triang}), fits of expression (\ref{AvCoop}) to the whole data on $p_{T}$ spectra for the most central class of Pb-Pb collisions at $\sqrt{s_{NN}} = 2.76$ TeV \cite{Abelev:2012wca} resulted in $\chi^{2}/n_{dof} = 1.49$ with $\emph{p-value} = 2 \cdot  10^{-6}$ ($n_{dof}=234$), which is still unacceptable.

\begin{figure}
\resizebox{0.46\textwidth}{!}{%
 \includegraphics{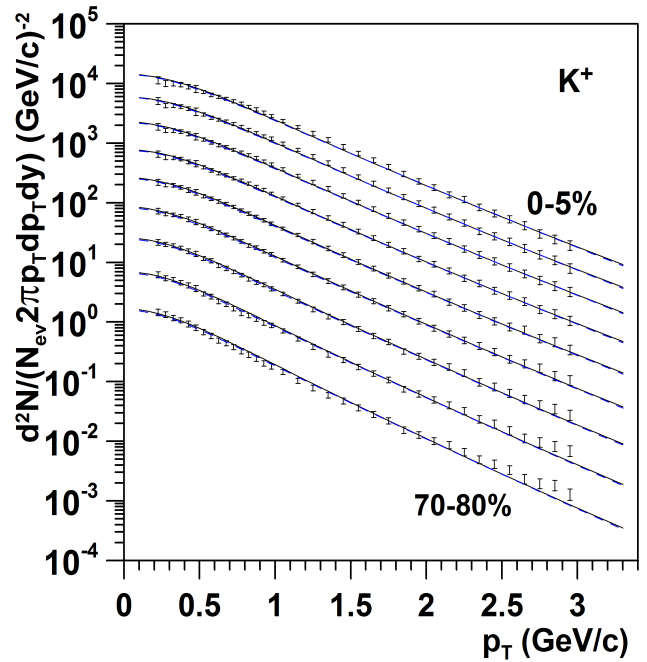}
}
\caption{Spectra of positive kaons measured in Pb-Pb collisions at $\sqrt{s_{NN}} = 2.76$ TeV, data used in the fit are presented as error bars only \protect\cite{Abelev:2013vea}, errors are sums of statistical and systematic components added in quadrature. Central to peripheral data are shown, spectra are scaled by factors $2^n$ (peripheral data not scaled). Lines are fits for the log-normal p.d.f. of $\beta_f$, dashed lines (blue) show fits of the SFOM without randomization and with all hadronic resonances included. }
\label{fig:3}
\end{figure}

\begin{figure}
\resizebox{0.46\textwidth}{!}{%
 \includegraphics{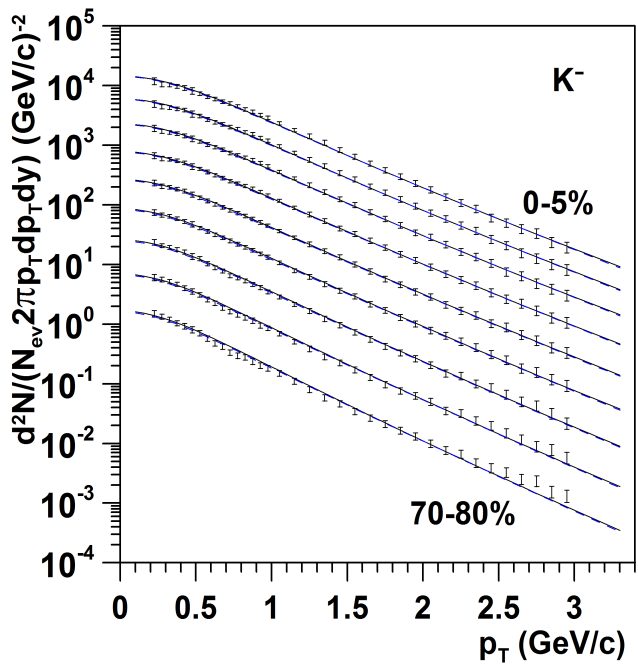}
}
\caption{The same as Fig.\,\ref{fig:3} but for negative kaons. }
\label{fig:4}
\end{figure}

The second assumption of the model is purely \textit{heuristic} - it states that the most weakly bound resonances should be removed from the hadron gas. To be more precise, all resonances with the full width $\Gamma > 250$ MeV (and masses below 1600 MeV) are removed \cite{Agashe:2014kda}. These are: $f_0(500)$, $h_1(1170)$, $a_1(1260)$, $\pi(1300)$, $f_0(1370)$, $\pi_1(1400)$, $a_0(1450)$, $\rho(1450)$, $K^*_0(1430)$ and $N(1440)$ (see footnote\footnote{In fact, the hint for this assumption was the accidental observation that after the update of the $f_0(500)$ mass to the lower one \cite{Agashe:2014kda}, the quality of the fit became worse. \label{przyp2}}). It should be noticed that the note attached to $f_0(500)$ says: ''The interpretation of this entry as a particle is controversial'' \cite{Agashe:2014kda} and the removal of this resonance has found a theoretical justification recently \cite{Broniowski:2015oha}. The exclusion of only $f_0(500)$
moves fits to the boundary of the acceptance, $\chi^{2}/n_{dof} \sim 1.3$ ( $\emph{p-value} \sim 0.001$), nevertheless according to the rigorous rules of the statistical inference it is still not a ''good'' fit \cite{Cowan:1998ji}. The removed resonances are weakly bound already in the vacuum, with the average lifetime $\tau < 1$ fm, so it might happen that they are not formed in the hot and dense medium at all, at least in the case of central Pb-Pb collisions at extreme energy $\sqrt{s_{NN}} = 2.76$ TeV. Precisely, resonances correspond to attractive interactions between hadrons. In medium, this interactions are likely modified and one cannot exclude the possibility that they might be weakened to such an extend that some resonances disappear before the freeze-out already. Anyway, this is a \textit{heuristic} hypothesis, but it works very well. It should be stressed at this point that both assumptions are necessary, if only the removal of weakly bound resonances is applied (no randomization of any parameter), the fit for the most central class is still unacceptable, $\chi^{2}/n_{dof} = 1.5$ ($\emph{p-value} =  10^{-6}$). It looks like both assumptions (phenomena) strengthen each other.

\section{Results}
\label{result}

\begin{figure}
\resizebox{0.46\textwidth}{!}{%
 \includegraphics{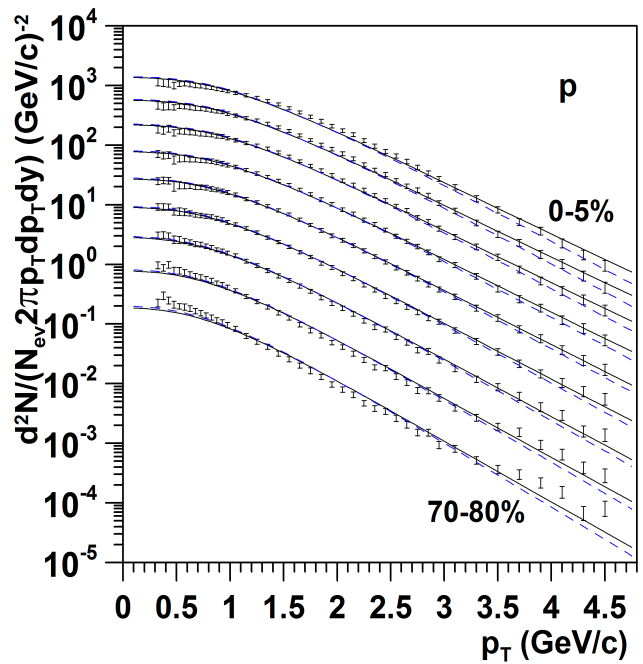}
}
\caption{Spectra of protons measured in Pb-Pb collisions at $\sqrt{s_{NN}} = 2.76$ TeV, data used in the fit are presented as error bars only \protect\cite{Abelev:2013vea}, errors are sums of statistical and systematic components added in quadrature. Central to peripheral data are shown, spectra are scaled by factors $2^n$ (peripheral data not scaled). Lines are fits for the log-normal p.d.f. of $\beta_f$, dashed lines (blue) show fits of the SFOM without randomization and with all hadronic resonances included. }
\label{fig:5}
\end{figure}

\begin{figure}
\resizebox{0.46\textwidth}{!}{%
 \includegraphics{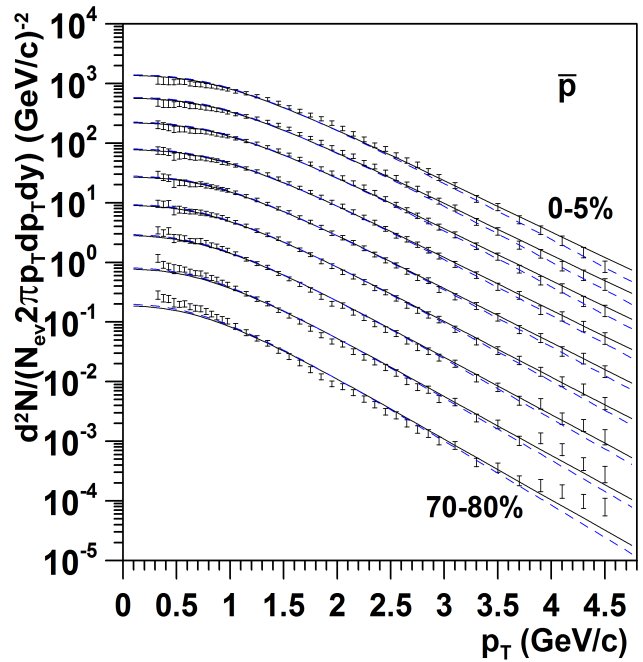}
}
\caption{The same as Fig.\,\ref{fig:5} but for antiprotons. }
\label{fig:6}
\end{figure}

\begin{figure}
\resizebox{0.46\textwidth}{!}{%
 \includegraphics{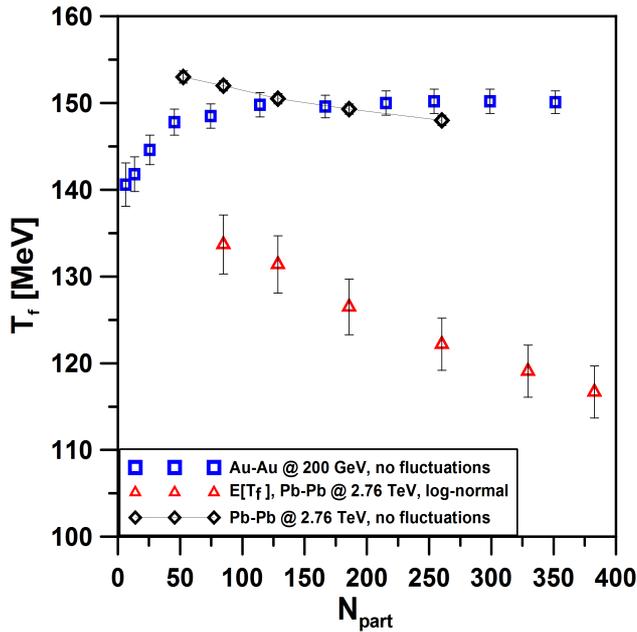}
}
\caption{Centrality dependence of the freeze-out temperature, Au-Au results are from \protect\cite{Prorok:2007xp}.}
\label{fig:7}
\end{figure}

The results of fits are presented in Tables~\ref{Table1}-\ref{Table3} and depicted in Figs.\,\ref{fig:1}-\ref{fig:6}. It has turned out that fits for both p.d.f.'s, Eqs.~(\ref{lognorm}) and (\ref{triang}), are the same practically, i.e. they are not distinguishable in figures, so only results for the log-normal p.d.f. are plotted. Fits of the original SFOM (with all hadronic resonances included) are also depicted in Figs.\,\ref{fig:1}-\ref{fig:6}.

In the RSFOM the production of low-$p_T$ pions is enhanced slightly in comparison with the results of the SFOM in central bins. For higher $p_T$ in central bins and for all other bins fits of pions are the same in both models. Fits of kaons are practically the same and both models underestimate high-$p_T$ production in peripheral bins. Results for low-$p_T$ protons and antiprotons are practically the same, slight overestimation but within errors for central bins goes gradually,  with the centrality deterioration, to underestimation in peripheral bins. In high-$p_T$ region ($p_T > 3$ GeV/c) fits of the RSFOM and the SFOM disagree and the disagreement deepens with $p_T$. But both fits agree with the data within errors for first 5 centrality bins starting from the most central one. For higher centrality classes fits underestimate the high-$p_T$ production (the SFOM more).

\begin{table*}
\caption{ Fit results for Pb-Pb collisions at $\sqrt{s_{NN}} = 2.76$ TeV and the measurement at central rapidity, $\mid y \mid < 0.5$, $n_{dof}=234$. Parameters of the log-normal p.d.f. have no units, so their values correspond to $\beta_f$ counted in MeV$^{-1}$ implicitly. } \label{Table1}
\begin{center}
\begin{tabular}{lllllllll} \hline\noalign{\smallskip}
 \multicolumn{9}{c}{log-normal p.d.f.}
\\
 \hline\noalign{\smallskip}
 Centrality & $\tau_f$ & $\rho_{max}$ & $\mu$ & $\sigma$ & $E[T_f]$ & $\sqrt{V[T_f]}$ & $\chi^{2}/n_{dof}$ & $\emph{p-value}$
\\
 & $(\textrm{fm})$ & $(\textrm{fm})$ &  &  & (MeV) & (MeV) &  & $(\%)$
\\
\noalign{\smallskip}\hline\noalign{\smallskip}
 0-5\% & 13.80 $\pm$ 0.40 & 20.48 $\pm$ 0.60 & -4.7439 $\pm$ 0.0235 & 0.1764 $\pm$ 0.0090 & 116.7 $\pm$ 3.0 & 20.7 $\pm$ 1.6 & 1.048 & 29
\\ 5-10\% & 12.60 $\pm$ 0.36 & 18.71 $\pm$ 0.54 & -4.7654 $\pm$ 0.0233 & 0.1687 $\pm$ 0.0093 & 119.1 $\pm$ 3.0 & 20.2 $\pm$ 1.7 & 0.844 & 96
\\ 10-20\% & 11.06 $\pm$ 0.31 & 16.35 $\pm$ 0.47 & -4.7932 $\pm$ 0.0230 & 0.1595 $\pm$ 0.0096 & 122.2 $\pm$ 3.0 & 19.6 $\pm$ 1.7 & 0.589 & 100
\\ 20-30\% & 9.30 $\pm$ 0.27 & 13.57 $\pm$ 0.40 & -4.8294 $\pm$ 0.0232 & 0.1467 $\pm$ 0.0105 & 126.5 $\pm$ 3.2 & 18.7 $\pm$ 1.9 & 0.335 & 100
\\ 30-40\% & 7.76 $\pm$ 0.23 & 11.04 $\pm$ 0.31 & -4.8694 $\pm$ 0.0232 & 0.1308 $\pm$ 0.0117 & 131.4 $\pm$ 3.3 & 17.3 $\pm$ 2.0 & 0.302 & 100
\\ 40-50\% & 6.55 $\pm$ 0.20 & 9.01 $\pm$ 0.28 & -4.8878 $\pm$ 0.0237 & 0.1244 $\pm$ 0.0125 & 133.7 $\pm$ 3.4 & 16.7 $\pm$ 2.2 & 0.609 & 100
\\ 50-60\% & 5.49 $\pm$ 0.18 & 7.23 $\pm$ 0.24 & -4.8957 $\pm$ 0.0245 & 0.1236 $\pm$ 0.0130 & 134.7 $\pm$ 3.6 & 16.7 $\pm$ 2.3 & 1.373 & 0.013
\\ 60-70\% & 4.56 $\pm$ 0.16 & 5.75 $\pm$ 0.21 & -4.8852 $\pm$ 0.0248 & 0.1298 $\pm$ 0.0125 & 133.4 $\pm$ 3.6 & 17.4 $\pm$ 2.2 & 2.821 & 0
\\ 70-80\% & 3.64 $\pm$ 0.14 & 4.41 $\pm$ 0.17 & -4.8775 $\pm$ 0.0253 & 0.1323 $\pm$ 0.0124 & 132.5 $\pm$ 3.6 & 17.6 $\pm$ 2.2 & 4.360 & 0
\\
\noalign{\smallskip}\hline
\end{tabular}
\end{center}
\end{table*}

\begin{table*}
\caption{ Fit results for Pb-Pb collisions at $\sqrt{s_{NN}} = 2.76$ TeV and the measurement at central rapidity, $\mid y \mid < 0.5$, $n_{dof}=234$. }\label{Table2}
\begin{center}
\begin{tabular}{lllllllll} \hline\noalign{\smallskip}
 \multicolumn{9}{c}{triangular p.d.f.}
\\
 \hline\noalign{\smallskip}
 Centrality & $\tau_f$ & $\rho_{max}$ & $\breve{\beta}_f$ & $\Gamma$ & $E[T_f]$ & $\sqrt{V[T_f]}$ & $\chi^{2}/n_{dof}$ & $\emph{p-value}$
\\
  & $(\textrm{fm})$ & $(\textrm{fm})$ & (MeV$^{-1}$) & (MeV$^{-1}$) & (MeV) & (MeV) & & $(\%)$
\\
\noalign{\smallskip}\hline\noalign{\smallskip}
 0-5\% & 14.42 & 21.45 & 0.0092482 & 0.0040906 & 111.6 & 22.6 & 1.026 & 38
\\ 5-10\% & 12.26 & 18.21 & 0.0084459 & 0.0031706 & 121.0 & 20.7 & 0.843 & 96
\\ 10-20\% & 10.68 & 15.86 & 0.0082155 & 0.0029280 & 124.1 & 20.2 & 0.602 & 100
\\ 20-30\% & 8.99 & 13.13 & 0.0078515 & 0.0025029 & 129.2 & 18.9 & 0.338 & 100
\\ 30-40\% & 7.54 & 10.74 & 0.0075161 & 0.0020650 & 134.4 & 17.2 & 0.309 & 100
\\ 40-50\% & 6.17 & 8.49 & 0.0072164 & 0.0016613 & 139.4 & 15.4 & 0.624 & 100
\\ 50-60\% & 5.05 & 6.66 & 0.0070426 & 0.0014141 & 142.5 & 14.3 & 1.396 & 0.006
\\ 60-70\% & 4.16 & 5.26 & 0.0070347 & 0.0014203 & 142.7 & 14.3 & 2.848 & 0
\\ 70-80\% & 3.25 & 3.94 & 0.0069937 & 0.0013403 & 143.4 & 13.9 & 4.389 & 0
\\
\noalign{\smallskip}\hline
\end{tabular}
\end{center}
\end{table*}

In the most central classes, where only the RSFOM works, the determined temperature is of the order of 110-120 MeV, which is much lower than the estimate from yields, $T_{ch} \simeq 156$ MeV \cite{Stachel:2013zma} but agrees qualitatively with the values of the kinetic freeze-out temperature given in \cite{Melo:2015wpa} and based on the blast-wave model \cite{Schnedermann:1993ws} fits.

In the mid-central region both approaches, i.e. the RSFOM and the SFOM, give acceptable fits, see Tables~\ref{Table1}-\ref{Table2} and Table~\ref{Table3}. This exactly means that both models cannot be rejected there. Applying the Ockham razor principle one should choose the simpler model in this case, that is the SFOM. One should also remember that values of the freeze-out temperature presented in Tables~\ref{Table1}-\ref{Table2} are the average values (over the sample), whereas the values of $T_f$ given in Table~\ref{Table3} (the case with the non-random freeze-out temperature) and in Table~\ref{Table4} (the same case but for Au-Au collisions at RHIC) are temperatures of "an average event" - one for each centrality class. One should notice here, that such "average event" might not have a real representative in the sample. Therefor the freeze-out temperatures from Tables~\ref{Table1}-\ref{Table2} and Table~\ref{Table3} are hardly similar and there is no reason they should be.

\section{Conclusions}
\label{conclus}

In summary, the chemical equilibrium Randomized Single-Freeze-Out Model has been applied successfully to the description of the production of identified hadrons measured at midrapidity in Pb-Pb collisions at $\sqrt{s_{NN}} = 2.76$ TeV \cite{Abelev:2013vea}. This has been achieved with the help of the more general, direct interpretation of the data and the removal of the most weakly bound resonances from the hadron gas. Additionally, the chemical equilibrium SFOM without the above-mentioned two new assumptions was examined in this context. The correct description of spectra measured at mid-central classes of Pb-Pb collisions at $\sqrt{s_{NN}} = 2.76$ TeV and the failure of the SFOM in the two most central classes might suggest new phenomena occurring there. These phenomena seem to appear at the two levels: in individual events, where the production of identified hadrons in each collision can be describe within the chemical equilibrium SFOM but with the reduced content of the hadron gas, and in the whole sample, causing substantial differences among collisions belonging to the same central class. As a result, the two most central bins of Pb-Pb collisions at $\sqrt{s_{NN}} = 2.76$ TeV seem to be significantly inhomogeneous, during each event the thermal system is created indeed and with approximately the same size at its end, however with different temperature. The distribution of the freeze-out temperature means the distribution within a bin here. But the significant part of the freeze-out temperature fluctuations might be of \textit{non-thermal} origin, so this would represent the possible variation of the freeze-out conditions event-by-event within the bin. And the final shape of the spectra is the consequence of summing emissions from many different sources.


In conclusion, the centrality bins of Pb-Pb collisions at $\sqrt{s_{NN}} = 2.76$ TeV can be divided into 3 groups: the first, the 2 most central bins where the freeze-out temperature fluctuates significantly; the second, the mid central bins where the situation looks similar to that at the RHIC, the same freeze-out temperature, $T_f \sim 150$ MeV (see Fig.\,\ref{fig:7}), only $\rho_{max}$ factor $\sim 1.5$ greater ($\tau_f$ approx. the same) which causes that the volume is greater $\sim 2.5$ times; the third, the peripheral bins where both approaches failed.

\begin{table*}
\caption{ Results for Pb-Pb collisions at $\sqrt{s_{NN}} = 2.76$ TeV and the measurement at central rapidity, $\mid y \mid < 0.5$, $n_{dof}=235$. Fits are for the SFOM without randomization and with all hadronic resonances included.}\label{Table3}
\begin{center}
\begin{tabular}{llllll} \hline\noalign{\smallskip}
 Centrality & $\tau_f$ $(\textrm{fm})$ & $\rho_{max}$ $(\textrm{fm})$ & $T_f$ (MeV) & $\chi^{2}/n_{dof}$ & $\emph{p-value}\;(\%)$
\\
\noalign{\smallskip}\hline\noalign{\smallskip}
 0-5\% & 9.96 $\pm$ 0.09 & 14.67 $\pm$ 0.89 & 147.0 $\pm$ 0.5 & 1.739 & $1.6 \cdot 10^{-9}$
\\ 5-10\% & 9.32 $\pm$ 0.08 & 13.70 $\pm$ 0.84 & 147.3 $\pm$ 0.5 & 1.445 & $9.2 \cdot 10^{-4}$
\\ 10-20\% & 8.41 $\pm$ 0.08 & 12.34 $\pm$ 0.78 & 148.0 $\pm$ 0.6 & 1.087 & 17.2
\\ 20-30\% & 7.29 $\pm$ 0.07 & 10.55 $\pm$ 0.70 & 149.3 $\pm$ 0.6 & 0.685 & 99.99
\\ 30-40\% & 6.30 $\pm$ 0.07 & 8.89 $\pm$ 0.62 & 150.5 $\pm$ 0.6 & 0.486 & 100
\\ 40-50\% & 5.36 $\pm$ 0.07 & 7.30 $\pm$ 0.54 & 152.0 $\pm$ 0.6 & 0.658 & 99.999
\\ 50-60\% & 4.48 $\pm$ 0.07 & 5.85 $\pm$ 0.48 & 153.0 $\pm$ 0.7 & 1.244 & 0.65
\\ 60-70\% & 3.70 $\pm$ 0.07 & 4.61 $\pm$ 0.41 & 153.5 $\pm$ 0.7 & 2.471 & 0
\\ 70-80\% & 2.89 $\pm$ 0.07 & 3.44 $\pm$ 0.34 & 154.1 $\pm$ 0.8 & 3.856 & 0
\\
\noalign{\smallskip}\hline
\end{tabular}
\end{center}
\end{table*}

\begin{table*}
\caption{ Results for PHENIX Au-Au collisions at $\sqrt{s_{NN}}=200$ GeV and the measurement at central rapidity, $\mid y \mid < 0.35$, $n_{dof}=122$, from \protect\cite{Prorok:2007xp}. Fits are for the SFOM without randomization and with all hadronic resonances included.}\label{Table4}
\begin{center}
\begin{tabular}{llllll} \hline\noalign{\smallskip}
 Centrality & $\tau_f$ $(\textrm{fm})$ & $\rho_{max}$ $(\textrm{fm})$ & $T_f$ (MeV) & $\mu_{B}$ (MeV) & $\chi^{2}/n_{dof}$
\\
\noalign{\smallskip}\hline\noalign{\smallskip}
 0-5\% & 9.5 $\pm$0.2 & 9.3$\pm$0.2 & 150.1$\pm$1.3 & 24.1$\pm$3.7 & 0.69
\\
5-10\% & 8.8$\pm$0.2 & 8.8$\pm$0.2 & 150.2$\pm$1.4 & 23.5$\pm$3.7 & 0.50
\\
10-15\% & 8.2$\pm$0.2 & 8.3$\pm$0.2 & 150.2$\pm$1.4 & 22.8$\pm$3.7 & 0.37
\\
15-20\% & 7.7$\pm$0.2 & 7.8$\pm$0.2 & 150.0$\pm$1.4 & 22.4$\pm$3.7 & 0.37
\\
20-30\% & 7.0$\pm$0.1 & 7.1$\pm$0.2 & 149.6$\pm$1.3 & 24.0$\pm$3.5 & 0.45
\\
30-40\% & 6.0$\pm$0.1 & 6.1$\pm$0.1 & 149.8$\pm$1.4 & 23.8$\pm$3.6 & 0.66
\\
40-50\% & 5.3$\pm$0.1 & 5.3$\pm$0.1 & 148.5$\pm$1.4 & 22.5$\pm$3.7 & 0.89
\\
50-60\% & 4.6$\pm$0.1 & 4.4$\pm$0.1 & 147.8$\pm$1.5 & 22.0$\pm$4.0 & 0.96
\\
60-70\% & 3.9$\pm$0.1 & 3.6$\pm$0.1 & 144.6$\pm$1.7 & 21.6$\pm$4.6 & 1.12
\\
70-80\% & 3.2$\pm$0.1 & 2.8$\pm$0.1 & 141.8$\pm$2.0 & 24.1$\pm$5.7 & 1.23
\\
80-92\% & 2.8$\pm$0.1 & 2.2$\pm$0.1 & 140.6$\pm$2.5 & 14.3$\pm$7.1 & 1.13
\\
\noalign{\smallskip}\hline
\end{tabular}
\end{center}
\end{table*}


And last, but not least, a great deal of data in high energy physics are averages, so in any theoretical modeling (of these data) one should be aware of possible misinterpretations when an average is compared with a prediction for a single event.



\begin{acknowledgement}
This work was not supported by any financial grant. Most of the calculations have been carried out using resources provided by Wroclaw Centre for Networking and Supercomputing \newline (http://wcss.pl), grant No. 268.
\end{acknowledgement}

%

%

%

\end{document}